# ARITHMETIC OPERATIONS IN MULTI-VALUED LOGIC

Vasundara Patel k s[1], k s gurumurthy[2]

[1]*Dept of ECE, BMSCE, Vishweshwaraiah Technological University,, Bangalore,*
vasundara.rs@gmail.com
[2]*Dept of E&C, UVCE, Bangalore*
drksgurumurthy@gmail.com

## *Abstract*

*This paper presents arithmetic operations like addition, subtraction and multiplications in Modulo-4 arithmetic, and also addition, multiplication in Galois field, using multi-valued logic (MVL). Quaternary to binary and binary to quaternary converters are designed using down literal circuits. Negation in modular arithmetic is designed with only one gate. Logic design of each operation is achieved by reducing the terms using Karnaugh diagrams, keeping minimum number of gates and depth of net in to consideration. Quaternary multiplier circuit is proposed to achieve required optimization. Simulation result of each operation is shown separately using Hspice.*

## 1. INTRODUCTION

Over the last three decades, designs using multiple-valued logic have been receiving considerable attention [1]. The history of Multiple-valued logic (MVL) as a separate subject began in the early 1920 by a polish philosopher Lukasiewicz. His intention was to introduce a third additional value to binary. The outcome of this investigation is known the Lukasiewicz system. Parallel to this approach the American mathematician Emil Post introduced multiple-valued algebra known as post algebra.

In modern SOC design, the interconnection is becoming a major problem because of the bus width. This problem can be solved by using Multiple-valued logic interconnection. For example a conventional 16 - bit bus (0 and 1) represents 65536 combinations. If we code the output with Quaternary logic (0, 1, 2 and 3), the width of the bus is reduced from 16 to 8. As a result, we can reduce power and area requirement for the interconnection.

Down literal circuit (DLC) is one of the most useful circuit element in multi-valued logic. The DLC can divide the multi-valued signal into a binary state at an arbitrary threshold.

Quaternary signals are converted to binary signals before performing arithmetic operations. Results of arithmetic operations are also binary signals. Hence these binary signals are to be converted to quaternary signals. In this paper, quaternary to binary and binary to quaternary converter are designed, and also Modulo-4 arithmetic operations are performed in such a way to get minimum number of gates and minimum depth of net

### 1.1. Modular arithmetic

Modular arithmetic is the arithmetic of congruence's, sometimes known informally as "clock arithmetic." In modular arithmetic, numbers "wrap around" upon reaching a given fixed quantity, which is known as the modulus. Modular arithmetic can be handled mathematically by introducing a congruence relation on the integers that is compatible with the operations of the ring of integers: addition, subtraction, and multiplication. Modular arithmetic is referenced in number theory, group theory, ring theory, knot theory, abstract algebra, cryptography, computer science, chemistry and the visual and musical arts. In computer science, modular arithmetic is often applied in bitwise operations and other operations involving fixed-width, cyclic data structures. The modulo operation, as implemented in many programming languages and calculators, is an application of modular arithmetic that is often used in this context. work that has been carried out in this field is listed below.

A 32 X 32 bit Multiplier using Multiple-valued Current Mode Circuit has been fabricated in 2 –um CMOS technology. The implemented 32x32 bit multiplier based on the radix-4 signed digit number system is superior to the fastest CMOS binary multiplier reported [2]. Novel quaternary half adder, full adder, and a carry-look ahead adder are introduced by M Thoidis. In his paper, the proposed circuits are static and operate in voltage mode. There is no current flow in steady states and thus no static power dissipation [3]. Ricardo designed a new truly full adder quaternary circuit using 3 power supply lines and multi-Vt transistors. Proposed technique benefits large scale circuits since the much power dissipation with increased speed can lead to the development of extremely low energy circuits while sustaining the high performance required for many applications [4].

Quaternary full adders based on output generator sharing are proposed by Hirokatsu with reduction in delay and power [5].

### 1.2. Galois field

A field is an algebraic structure in which operations of addition, subtraction, multiplication, and division (except by zero) can be performed, and satisfy the usual rules. Field with finite number of elements is called Galois field. GF (2) is binary field and it can be extended to GF ($2^k$). These two fields are most widely used in digital data transmission and storage system [6].

GF ($2^k$) plays an important role in communications including error correcting codes, cryptography, switching and digital signal processing. In these applications, area and speed requirements of an IC are essential. Therefore an efficient hardware structure for such operations is desirable. Efficiency of these applications heavily depends on the efficiency of arithmetic operations in Galois fields like addition, multiplication, subtraction, inversion etc. Many of the private and public-key algorithms in cryptography aim to achieve high level security, which relies on computations in GF ($2^k$). Hence effective algorithm must be developed to carry out arithmetic operations in GF ($2^k$). Elements of GF ($2^k$) are 0 and 1. Most of the applications use GF ($2^k$) as their basis. Composite field GF (($2^n$)$^m$) is more suitable where k = nm. The field GF ($2^n$) over which the composite field is defined is called the ground field [7]. In the quaternary case, the ground field becomes $GF(2^2)$ = GF (4). Lot of work has been carried out in this field which is listed below

A new table look-up method for finding the log and antilog of finite field elements has been developed by N Glover. VLSI architecture is developed for his new algorithm to perform finite field arithmetic operations [8]. A new algorithm based on a pattern matching technique for computing multiplication and division of two elements in $GF(2^m)$ is developed by Kovac, M. Ranganathan and N.Varanasi [9]. An efficient VLSI architecture for implementing the proposed algorithm is described in this paper. An efficient architecture of a reconfigurable bit serial polynomial basis multiplier for Galois field GF ($2^m$) is explained in [10]. In public cryptosystems and error correcting over Galois field, $AB^2$ is an efficient basic operation [11].

This paper uses Multiple -valued logic approach to minimize the systolic architecture of algorithm over binary Galois fields.

A unique encoding technique to perform arithmetic operations in Galois field using Multiple-valued logic is presented in [7].The computations are done in quaternary logic system in this paper. A new inner product $AB^2$ multiplication algorithm and effective hardware architecture for exponentiation in finite fields $GF(2^m)$ is explained in [12].

Time-independent Montgomery multiplication algorithm is presented in [13]. These architectures are optimized to reduce silicon area and to reduce multiplication time delay.A pipelined Inversion and division circuit is designed in Galois field using $AB^2$ circuit technique[14]. This circuit shows significant amount of savings on both transistor count and connections.

## 2. DESIGN OF QUATERNARY CONVERTER CIRCUITS

### 2.1. Objective of optimization

Objective of optimization is to minimize number of gates needed and also to minimize depth of net. Depth of net is the largest number of gates in any path from input to output. The reason for choosing these two objectives is that they will give very good properties when implemented in VLSI. Minimizing number of gates will reduce the chip area, and minimizing depth will give opportunity to use highest clock frequency.

### 2.2. Quaternary to binary converter

A basic Quaternary to binary converter uses three down literal circuits DLC1, DLC2, DLC3 and 2:1 multiplexer as shown in figure 2. Q is the quaternary input varying as 0, 1, 2 and 3 which is given to three DLC circuits. The binary out puts thus obtained will be in complemented form and are required to pass through inverters to get actual binary numbers.

Down literal circuits are realized from basic CMOS inverter by changing the threshold voltages of pmos and nmos transistors as shown in figure 1. Truth table of DLC1, DLC2 and DLC3 is shown in table 1.

### 2.3. Binary to quaternary converter

Binary to quaternary converter circuit is shown in figure 3. LSB and MSB of a two bit binary number are given to DLC 1 as shown in the figure 3. Threshold voltages of transistor M1 = - 0.6 V, M2 = 0.6 V, M3 = - 1.2 V and M4 = 0.6 V. output of two inverters will provide quaternary number. DLC1: Vtp = -2.2V and Vtn = 0.2V , DLC2: Vtp = -1.2V and Vtn = 1.2V, DLC3: Vtp = 0.2V and Vtn = 2.2V.

Table 1: Truth table of down literal circuits

| In | Out | | |
|---|---|---|---|
|    | DLC1 | DLC2 | DLC3 |
| 0 | 3 | 3 | 3 |
| 1 | 0 | 3 | 3 |
| 2 | 0 | 0 | 3 |
| 3 | 0 | 0 | 0 |

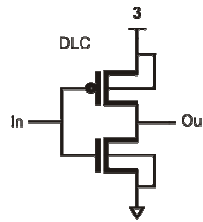

Figure 1: Circuit diagram for DLC

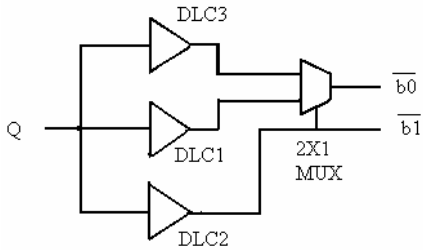
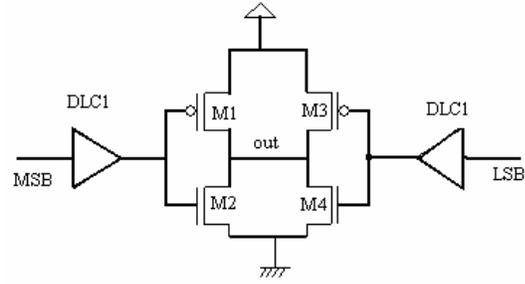

Figure 2: Quaternary to binary Converter

Figure 3: Quaternary to binary converter

## 3. MODULO-4 ARITHMETIC OPERATIONS

### 3.1. Modulo-4 addition and multiplication

Modulo-4 addition and multiplication is performed after quaternary to binary conversion. Quaternary inputs 0, 1, 2, 3 are represented in binary as 00, 01, 10, and 11 respectively [14]. Natural representation of quaternary numbers is shown in table 2. Quaternary inputs are converted to binary with the help of converter circuit shown in figure 2. Modulo-4 addition and multiplication are performed by taking objective of optimization in to consideration. Truth table of Modulo-4 addition and multiplication are shown in table 3. These tables show the quaternary numbers which are to be converted to binary before performing the addition and multiplication operations.

Minimal functions have been obtained from the Karnaugh diagrams for the addition and multiplication tables shown in table 3 and then simplified as much as possible using all possible gate types. Minimal functions obtained from the minimal polynomials extracted from the Karnaugh diagrams are shown below. Let $x_1 x_2$ and $y_1 y_2$ be the binary representation of quaternary numbers which has to be added and multiplied.

Let $m_1$ and $m_2$ denote the binary result of multiplying the binary numbers $x_1 x_2$ and $y_1 y_2$, and $a_1 a_2$ the result when adding them. Hence binary to quaternary converters are required to get the quaternary outputs.

For addition:

$a_1 = (x_1 \oplus y_1) \oplus (x_2 y_2)$

$a_2 = (x_2 \oplus y_2)$

For multiplication:

$m_1 = x_1 y_1' y_2 + x_1 x_2' y_2 + x_1' x_2 y_1 + x_2 y_1 y_2'$

$\phantom{m_1} = (x_1 y_2) \oplus (x_2 y_1)$

$m_2 = x_2 y_2$

Logical implementation for addition and multiplication in modulo-4 are shown in figure 4 and figure 5 respectively. From these two figures it is understood that depth of net is reduced to two and minimum number gates required are four.

Table 2: Natural representation of quaternary numbers.

| Quaternary logic | Natural Representation |
|---|---|
| 0 | 00 |
| 1 | 01 |
| 2 | 10 |
| 3 | 11 |

Table 3: Tables for modulo-4 addition and multiplication

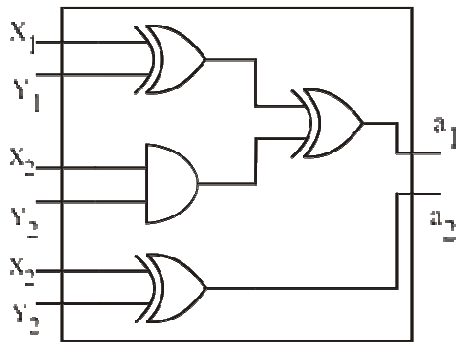

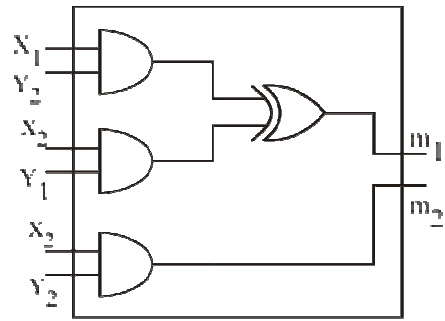

Figure 4: Logic diagram for modulo-4 addition

Figure 5: Logic diagram for modulo-4 multiplication

### 3.2. Modulo-4 subtraction, negation and quaternary number multiplied by 2

Modulo-N addition and multiplication are commutative in nature, whereas Modulo-N subtraction is not commutative. Hence we are subtracting $y_1 y_2$ from $x_1 x_2$. If X is the quaternary value of $x_1 x_2$ and Y is the quaternary value of $y_1 y_2$, then $S = (X + 4) - Y$, if $X < Y$. For example if $X = 2$ and $Y = 3$ then $2 + 4 = 6 - 3 = 3$, table 4 shows subtraction of $y_1 y_2$ from $x_1 x_2$.

International Journal of VLSI design & Communication Systems (VLSICS), Vol.1, No.1, March 2010

Table 4: Table for modulo-4 subtraction

|       |   | $y_1 y_2$ |   |   |
|-------|---|---|---|---|
| $x_1 x_2$ | 0 | 1 | 2 | 3 |
| 0     | 0 | 3 | 2 | 1 |
| 1     | 1 | 0 | 3 | 2 |
| 2     | 2 | 1 | 0 | 3 |
| 3     | 3 | 2 | 1 | 0 |

Minimal functions have been obtained from the Karnaugh diagrams for the subtraction between two quaternary numbers and then simplified as much as possible using all possible gate types. Minimal functions obtained from the minimal polynomials extracted from the Karnaugh diagrams are shown below. Let $x_1 x_2$ and $y_1 y_2$ be the binary representation of quaternary numbers which has to be subtracted.

Let $s_1$ and $s_2$ denote the binary result of subtracting the binary numbers $x_1 x_2$ and $y_1 y_2$.

$s_1 = ( x_1 \oplus y_1 ) \oplus ( x_2' y_2 )$

$s_2 = x_2 y_2' + x_2' y_2 = (x_2 \oplus y_2)$

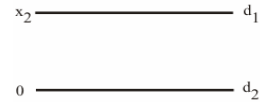

Logical implementation for subtraction in modulo-4 is shown in figure 6. The circuit is almost similar to addition and multiplication circuits.

Negation is always used in modular arithmetic. Negation of elements in Moudlo-4 arithmetic can be viewed as Multiplication by constant 3. Let x is the quaternary number which represents 0, 1, 2 and 3. $x_1 x_2$ are the binary numbers after conversion from quaternary. Let $n_1$ and $n_2$ denote the result of negating numbers $x_1 x_2$. It is observed from the table shown in figure 7 that $n_2$ and $x_2$ columns are similar. $n_1$ is the result of exor operation of $x_1$ and $x_2$. Hence we require one exor gate for negating X. Logic diagram for negation is shown in figure 7.

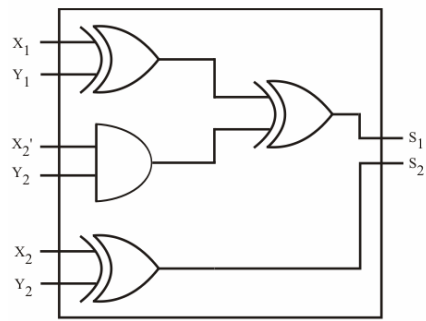

Figure 6: Logic diagram for modulo-4 subtraction

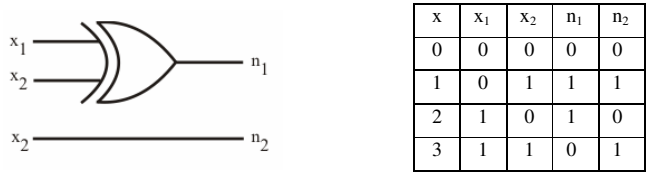

| x | $x_1$ | $x_2$ | $n_1$ | $n_2$ |
|---|---|---|---|---|
| 0 | 0 | 0 | 0 | 0 |
| 1 | 0 | 1 | 1 | 1 |
| 2 | 1 | 0 | 1 | 0 |
| 3 | 1 | 1 | 0 | 1 |

Figure 7: Logic diagram and truth table for negation

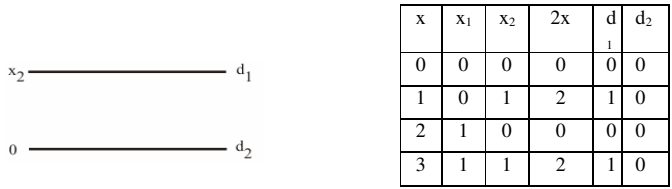

| x | $x_1$ | $x_2$ | 2x | $d_1$ | $d_2$ |
|---|---|---|---|---|---|
| 0 | 0 | 0 | 0 | 0 | 0 |
| 1 | 0 | 1 | 2 | 1 | 0 |
| 2 | 1 | 0 | 0 | 0 | 0 |
| 3 | 1 | 1 | 2 | 1 | 0 |

Figure 8: Logic diagram and Table for multiplication of x by 2

Multiplication of quaternary number x by number 2 is achived after converting x to binary number $x_1 x_2$. $d_1$ and $d_2$ are the result of multiplying $x_1 x_2$ by the number 2. Table in figure 8 shows that $d_1 = x_2$ and $d_2 = 0$. Logic diagram for the same is shown in figure 8.

## 4. GALOIS ADDITION AND MULTIPLICATION

The elements of GF (4) are 0, 1, 2, and 3, respectively, providing that 0 denotes the additive identity, and 1 denotes the multiplicative identity. Using these assumptions, the addition and multiplication operations in GF (4) can be defined as shown in figure 9 and figure 10 respectively. The two-bit binary representation of elements in GF (4) is shown in table 2.

### 4.1. Galois addition

Galois addition table in Figure 9 is used in Karnaugh diagrams to obtain minimum function. Minimal functions obtained from the minimal polynomials extracted from the Karnaugh diagrams for GF (4) addition is shown below. Let $x_1 x_2$ and $y_1 y_2$ be the binary representation of two quaternary numbers which have to be added. $a_1$ and $a_2$ are the two bit result of addition between $x_1 x_2$ and $y_1 y_2$.

$a_1 = (x_1 \oplus y_1)$

$a_2 = (x_2 \oplus y_2)$

Above equation shows that addition in GF (4) requires only two gates and depth of net is reduced to one. This is a very good design among four circuits. Logical implementation of the circuit is shown in figure 9.

### 4.2. Galois multiplication

Galois multiplication table between two quaternary numbers X and Y is shown in Figure 10. Quaternary numbers are converted to binary numbers and multiplication operation is performed in usual way to get binary results. These results are used in Karnaugh diagrams to obtain minimum function. Minimal functions obtained from the minimal polynomials extracted from the Karnaugh diagrams for GF (4) multiplication are shown below. Let $x_1 x_2$ and $y_1 y_2$ be the binary representation of two quaternary numbers which have to be multiplied. $m_1$ and m2 are result of multiplying $x_1 x_2$ and $y_1 y_2$.

$m_1 = x_1 y_1' y_2 + x_1 x_2' y_1 y_2' + x_1' x_2 y_1 + x_2 y_1 y_2$

$m_2 = x_1 y_1 \oplus x_2 y_2$

which shows the requirement of more number of gates for realizing $m_1$(Requires 54 transistors in addition to two converters, quaternary to binary and binary to quaternary). Hence we are proposing a new circuit, shown in figure 10. Signal needs to travel through maximum of 4 transistors. Proposed circuit gives regularity and reduced propagation delay. Quaternary signals need not be converted to binary signals. This quaternary multiplexer has three similar multiplexer blocks [16] and it works like binary multiplexer. X is the selection line for the main multiplexer. When the value on this line is 0, the first line (ground) is transmitted to output Q. When it is 1, the second line Y is transmitted. Third line is connected to a multiplexer with Y as select line. Quaternary numbers 0, 2, 3 and 1 are transmitted when X is 2. Similarly, when the select line X is 3, quaternary numbers 0, 3, 1 and 2 will be transmitted. Hence the table of Galois multiplier is verified

## 5. Simulations

HSPICE transient analysis simulation was done to verify the functionality of the circuits discussed in previous sections. Waveforms are analyzed on Cscope. TSMC 180nm technology files are used for the simulations.

Simulation result of quaternary to binary and binary to quaternary are shown in figure 11 and figure 12 respectively.

Simulation result of Modulo-4 addition is shown in figure 13. $a_1$ and $a_2$ are the result of polynomials shown in equations obtained by K map reduction from the addition table shown in table 3. Modulo-4 addition requires 40 transistors (4 gates) and multiplication requires 24 transistors (4 gates).Simulation result of Modulo-4 subtraction is shown in figure 14. $s_1$ and $s_3$ represents result of subtraction. Multiplication is shown in figure 15. $m_1$ and $m_2$ are the result of polynomials shown in equations obtained by K map reduction for the multiplication table shown in table 3.

Simulation result of Galois addition is shown in figure 16. $a_1$ and $a_2$ are the result of polynomials shown in equations obtained by K map reduction shown in addition table of figure 3. Galois addition requires only two gates and 24 transistors and can be reduced further. In Galois multiplication, quaternary inputs are multiplied directly which is suitable for required optimization. Simulation result of Quaternary multiplier is shown in figure 17. This is faster than other circuit and requires 72 transistors. Quaternary to binary converter at the input and binary to quaternary at the output are not required for this circuit where as other circuits require converters.

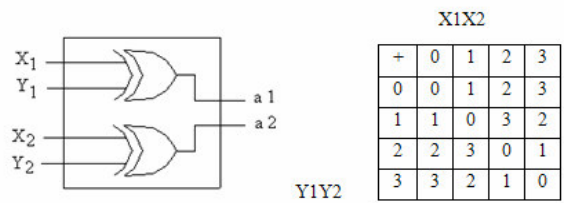

Figure 9: Logic diagram for GF adder and addition table for GF (4)

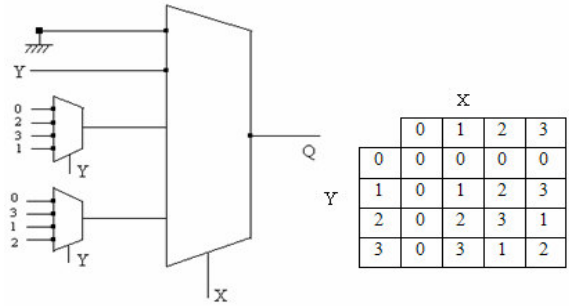

Figure 10: Logic diagram for multiplication in GF (4) and multiplication table for GF (4)

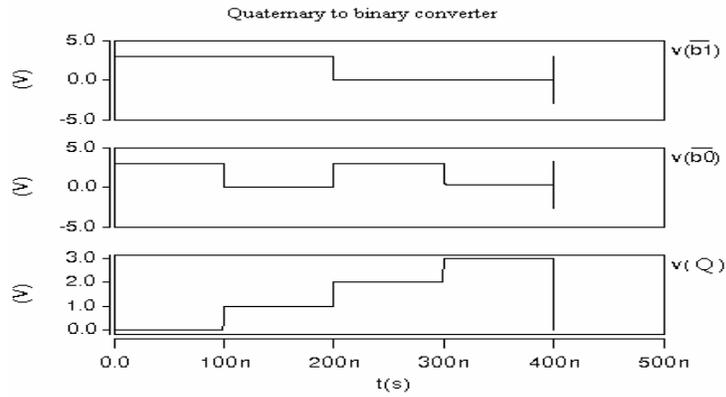

Figure 11: Simulatin result of quaternary to binary converter

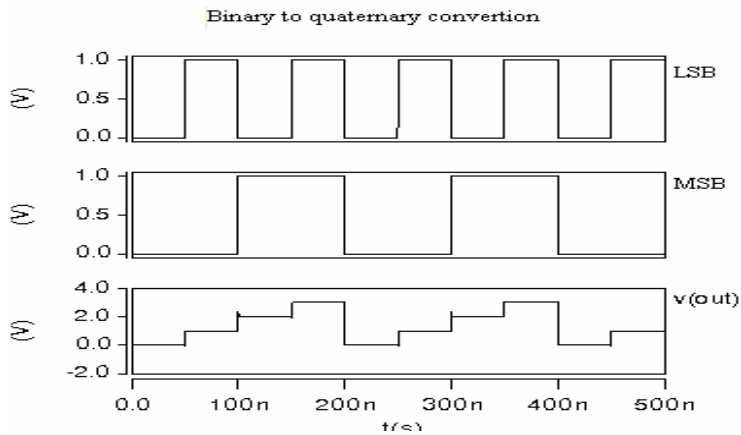

Figure 12: Simulatin result of binary to quaternary converter

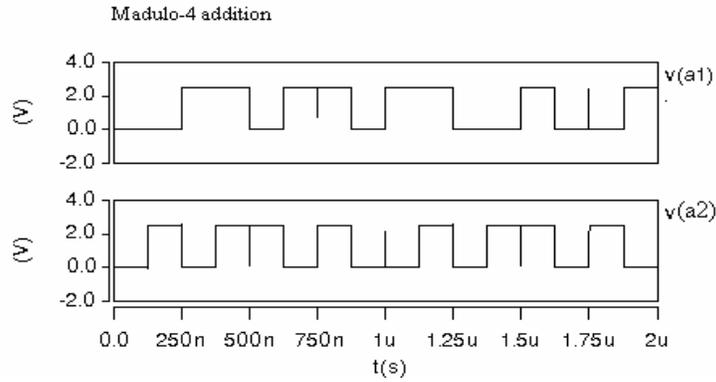

Figure 13: Simulatin result of addition in Modulo-4

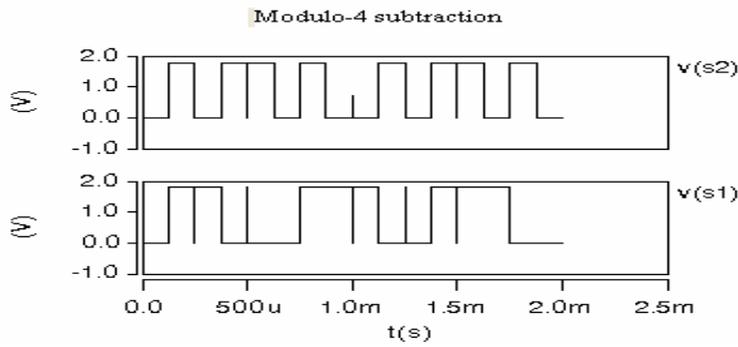

Figure 14: Simulation result of subtraction in Modulo-4

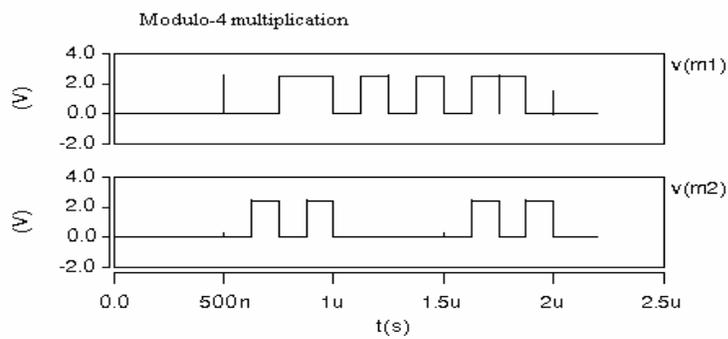

Figure 15: Simulation result of multiplication in Modulo-4

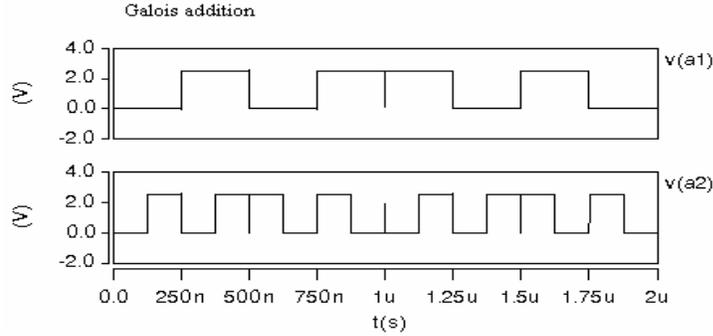

Figure 16: Simulatin result of Galois addition

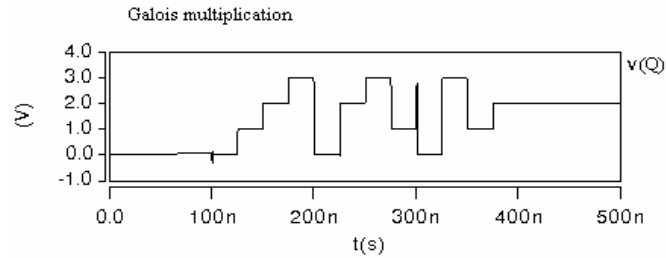

Figure 17: Simulatin result of Galois multiplication

## 6. CONCLUSIONS

Binary to quaternary and quaternary to binary converters are designed using down literal circuits. A quaternary Galois multiplier is proposed which uses three quaternary multiplexers and maintains regularity of the circuit. Implementation of the circuit shows higher performance than circuits using two variable representations. In [17] neuron MOSFETs are used, which consumes more power and more fabrication steps than conventional CMOS devices. Circuits for Modulo-4 addition, multiplication and subtraction require only 4 gates. Galois addition requires two xor gates which is most optimized one among other circuits while implementing in VLSI. With the help of quaternary logic levels, we have reduced the interconnections. We have also used less number of gates and hence less area for Galois and modulo-4 arithmetic operations. Proposed circuits are suitable for implementing in VLSI with less number of interconnections and less area.